# Improved superlensing in two-dimensional photonic crystals with a basis

X. Wang, Z.F. Ren and K. Kempa

Department of Physics

Boston College, Chestnut Hill, MA 02467

**Abstract**

We study propagation of light in square and hexagonal two-dimensional photonic crystals. We show, that slabs of these crystals focus light with subwavelength resolution. We propose a systematic way to increase this resolution, at an essentially fixed frequency, by employing a hierarchy of crystals of the same structure, and the same lattice constant, but with an increasingly complex basis.

wangxb@bc.edu

Pendry showed [1,2], that a slab made of a material with an isotropic refractive index $n = -1$, restores not only *phases* of the transmitted propagating waves, but also *amplitudes* of the evanescent waves that are responsible for the subwavelength details of the source geometry. Thus, such a material (metamaterial) can be used to make a superlens capable of the subwavelength imaging. Many proposals have been put forth in recent years to fabricate such a metamaterial. In one scheme [3-6], an effective medium made of metallic rods, and the so-called split-ring resonantors, has been shown to act as such a metamedium in the microwave frequency range. Other similar schemes have also been proposed [7,8].

In another approach, it has been shown that two-dimensional photonic crystals (2D-PCs) can act as effective media with a negative effective refractive index [9-11], however superlensisng has not been demonstrated. Recently, it was shown that the propagation of light in a square 2D-PC in the first photonic band leads to an all angle negative refraction (AANR), which leads to superlensing, but where an isotropic negative refractive index cannot be defined. It was also shown theoretically [12], that the resolution of the superlens is limited only by the coarsity of the PC. The negative refraction and the slab focusing in the second photonic band have recently been demonstrated experimentally in a square 2D-PC, again in the microwave frequency range [13].

In this letter we study propagation of light in square and hexagonal 2D-PCs, using the plane wave (PW) expansion, and the finite difference time domain (FDTD) techniques. We show, that the superlansing occurs in both systems, and propose a systematic way to increase its resolution at an essentially fixed frequency of the light.

This can be achieved in a hierarchy of crystals of the same structure, but with an increasingly complex basis.

The PCs studied here consist of a periodic (in the x-y plane) array of infinitely long, cylindrical (along z-direction) air-holes (or rods) of radius *r*, in a dielectric matrix. Each PC is assumed to have a finite width in the x-direction, but is unrestricted in the y direction. We obtain photonic band structures by numerically solving the Maxwell equations. We employ here the standard PW expansion [14]. To obtain maps of the propagating waves, we use the FDTD simulations with perfectly matched, layer boundary conditions [15-17].

We first study the *square* PC, which became a structure of choice in the recent experimental studies of the negative refraction. It was shown, that AANR can occur in this PC [12]. We choose PC of holes in a dielectric matrix, with parameters of Ref. [12], i.e. hole radius $r = 0.35a$, where the lattice constant is *a*, and the dielectric matrix with dielectric constant $\varepsilon = 12$. The calculated band structure (normalized frequency $\omega = a/\lambda$ vs. wave vector *k*) for the transverse electric (TE) mode (the magnetic field parallel to air holes) is shown in Fig. 1(a). The AANR occurs at the top of the first band (around the M-point) at the frequency $\omega = \omega_1$, indicated by the horizontal dashed line. The inset in Fig.1(a) shows the constant frequency contour obtained by crossing the first band with the equal energy surface for $\omega = \omega_1$. The fact that it is not circular, suggests a strong anisotropy of the wave propagation in this PC. In fact, a channeling of the wave propagation in the ΓM and equivalent directions occurs [18], due to larger group velocities of propagating waves in these directions.

Lets now consider the *hexagonal* 2D-PC of holes in a dielectric matrix, with parameters like those used in Ref. [9], i.e. $r = 0.4a$ and $\varepsilon = 12.96$. The photonic band structure for the transverse magnetic (TM) modes (the electric field parallel to cylindrical holes) is show in Fig. 1(b). The horizontal dashed line with frequency $\omega_2$ indicates the best frequency for the superlensing. The corresponding constant frequency contour is now essentially circular, the propagation of the waves isotropic, and one can define an isotropic effective refractive index. To illustrate this, we calculate the wave propagation maps. We turn-on at time $t_0$, a point source of cylindrical waves with frequency $\omega_2$, outside and to the left of the slab of the PC. Then, we calculate the field patterns at later times using the FDTD simulations. Fig. 2 shows the propagation maps for this case, for various positions of the point source, and thickness of the slab. It is easy to verify, that the propagation of light follows simple rules of the geometric optics with the Snell's-law refraction at each interface, and an effective isotropic refractive index $n = -1$. From the field intensity distribution across the image (approximately Gaussian), we find that the spatial resolution, defined as the ratio of the full width at half maximum (FWHM) of the image peak to the wavelength, is $R = 0.4$. This is smaller than the wavelength, and of the order of the inter-hole distance. Similar resolution can be obtained for the square crystal, however, because of the channeling effect discussed above, the lensing does not follow the rules of geometric optics. As expected, both crystals are capable of superlensing with a resolution limited by the coarsity of the crystal. We now propose a systematic scheme for increasing this subwavelength resolution.

It was shown [12], that the resolution of a superlens based on PC can be increased by reducing the normalized frequency at which the superlensing occurs. This is simply because reducing the normalized frequency means reducing the ratio of $a/\lambda$, which for a fixed $\lambda$ means reducing $a$, i.e. making the medium less coarse. Umklapp process is a well-known way of creating lower maxima in band structures. In general, doubling the size of a unit cell of a crystal, halves its Brillouin zone, and then an Umklapp process flips dispersion curves along the new zone edge creating "optical" branches with lower maxima.

In order to test this idea in the framework of a square PC, we consider an inverted structure in which air-holes are replaced with dielectric cylindrical rods. First, we consider a simple square (S) crystal sketched in Fig. 3(b) (upper panel). For this crystal we choose $r = 0.124a$, $\varepsilon = 9.2$. We calculate the photonic band structure for this crystal (TM modes). A section of this band structure in the ΓM direction is shown in Fig. 3(a) (solid line). We find that a slab of this crystal sharply images a point source. Electric field strength plotted along the image center, parallel to the slab, is shown in Fig.4 (solid line), and shows that the spatial resolution is $R = 0.5$. Next, we consider a crystal sketched in Fig.3(b) (lower panel). If $a$ was equal to $a'$, one could view this crystal as a modified S (MS) crystal, obtained by reducing the radius of every other rod in the S crystal (by about 30% in the calculations). A fragment of the normalized photonic band structure of this MS crystal is shown in Fig.3(b) (dashed line). As expected, the Umklapped part of the acoustic branch in the first band (optical branch) indeed develops a symmetric maximum at the Γ point, at the normalized frequency reduced by a factor of 40% relative to the corresponding maximum of the S crystal. Superlensing can occur in this modified crystal as well, at the normalized frequency $\omega' = a'/\lambda'$. By requiring that the MS crystal superlenses at the same absolute frequency (or wavelength) as the S crystal, i.e. $\lambda' = \lambda$, we find that $a' = a\omega'/\omega = 0.63a$. Clearly the MS crystal is less coarse and therefore its resolution should be better than that of S crystal. The corresponding electric field strength for the MS crystal, calculated along the image center, is shown in Fig. 4 (dashed line). The spatial resolution is now $R = 0.25$, by a factor of two better than that for the S crystal, confirming that $R\lambda$ is indeed roughly equal to the nearest neighbor inter-rod separation. Note, that the side maxima shown in Fig. 4 represent crests of propagating waves, and therefore they average-out to zero over time when calculating the light intensity. In contrast, the main image peak is stationary, and therefore relevant for the calculation of the resolution.

Another, alternative way of viewing this process of improving the spatial resolution of the imaging, without relying on the normalized frequency argument, is as follows. Notice, that the MS crystal sketched in Fig. 3(b) (lower panel) is also a square crystal, with about the same lattice constant, but with a basis, and turned about the axis vertical to the page by $45^0$. The basis is here the large-small rods pair. Therefore, by retaining the crystal structure, and the magnitude of the lattice constant, the main features of the band structure including the frequency range at which the superlansing occurs, are preserved. Complicating the basis, on the other hand, decreases the overall coarsity of the crystal, which, in turn, leads to the improved resolution. While these two schemes are equivalent in this example, the basis development scheme is more general, and can be applied to any crystal structure.

We now employ this scheme to generate a family of hexagonal PCs capable of superlansing with increasing resolution. The family consists of three structures (A,B and C), shown with their respective band structures in Fig.5. A is the hexagonal structure studied in the context of Figs. 1(b) and 2. Structures B and C are modifications of the A structure by addition of small extra holes, as shown in Fig. 5. B, C remain hexagonal, with the same lattice constant, but with a basis. The basis for structure B is the large hole (of structure A) plus one small hole. The basis for structure C is the large hole plus two small holes. As expected, the increased complexity of the basis does not affect the band structure significantly. In particular, the first inverted band where the frequency of the negative refraction lies, is essentially the same for all three structures. As a result, the superlensing occurs at approximately the same frequency, for all these structures. The propagation maps are shown in Fig. 6, and show that the slab imaging occurs for all the structures, and also that the sharpness of the image improves with increased complexity of the basis. This is confirmed by the electric field intensity calculated across the image (parallel to the edge of the crystal). The corresponding resolutions are 0.54, 0.43 and 0.27. Again, $R\lambda$ scales well with the nearest neighbor distance, for each structure.

In conclusion, we study propagation of light in square and hexagonal 2D-PCs. We show, that slabs of these crystals focus light with subwavelength resolution. We propose a systematic way of increasing resolution of this superlensing, at an essentially fixed frequency, by employing a hierarchy of photonic crystals of the same lattice structure, and lattice constant, but with an increasingly complex basis. We propose this as an efficient scheme of increasing resolution of the future PC superlenses.

This work is supported by US Army Research Development and Engineering Command, Natick Soldier Center, under the grant DAAD16-03-C-0052.

**Figure Captions**

**Fig. 1. Calculated photonic band structures for (a) square 2DPC with air-holes of radius $r = 0.35a$, in a dielectric matrix with $\varepsilon = 12$, and (b) hexagonal 2DPC with air-holes of radius $r = 0.4a$, in a dielectric matrix with $\varepsilon = 12.96$.**

**Fig. 2. The propagation map (electric field distribution across space) for a slab of the hexagonal 2D-PC (the corresponding band structure in Fig. 1(b)), for varying source positions and thickness of the slab. The point source frequency is $\omega_2 = 0.305$. Positions of the images follow the geometric optics analysis, in which the PC is considered a medium with $n = -1$, and a Snell's-law refraction occurs at each interface.**

**Fig. 3. Systematic scheme for improved image resolution via period doubling. (a) Photonic band structures of two 2DPC. Solid line is for the simple (S) square 2D-PC (sketched in b, upper panel), and dashed line is for the modified simple (MS) 2D-PC (sketched in b, lower panel).**

**Fig. 4. Light intensity scan through the image of a point source, parallel to the slab edge, for the S crystal (solid line), and for the MS crystal (dashed line). The increased resolution for the later case is clearly visible.**

**Fig. 5. Systematic scheme for improved image resolution via basis development. Photonic band gap structures (left) and the corresponding 2DPC crystal structures**

(right). Simple hexagonal structure A (upper panel), modified hexagonal structure with a simple basis consisting of two holes B (middle panel), and modified hexagonal structure with a more complicated basis consisting of three holes C (lower panel).

Fig. 6. Propagation maps for slabs of the hexagonal 2DPC family of structures, showing the superlensing capability of all structures. Structure A (upper panel), B (middle panel), and C (lower panel).

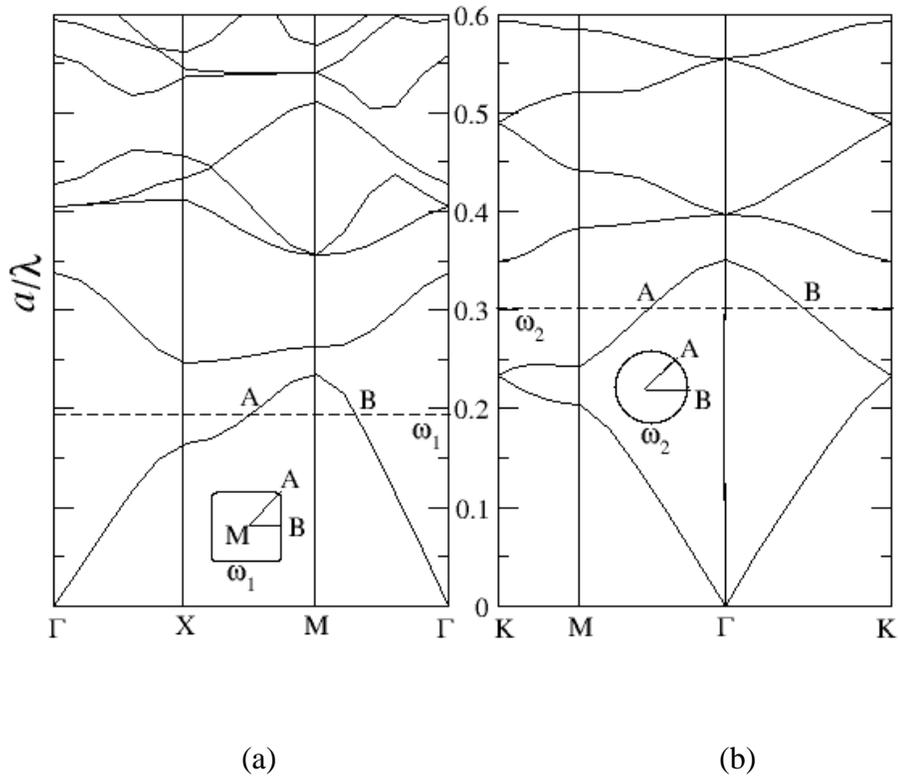

(a)                  (b)

**Fig. 1**

**X. Wang et al.**

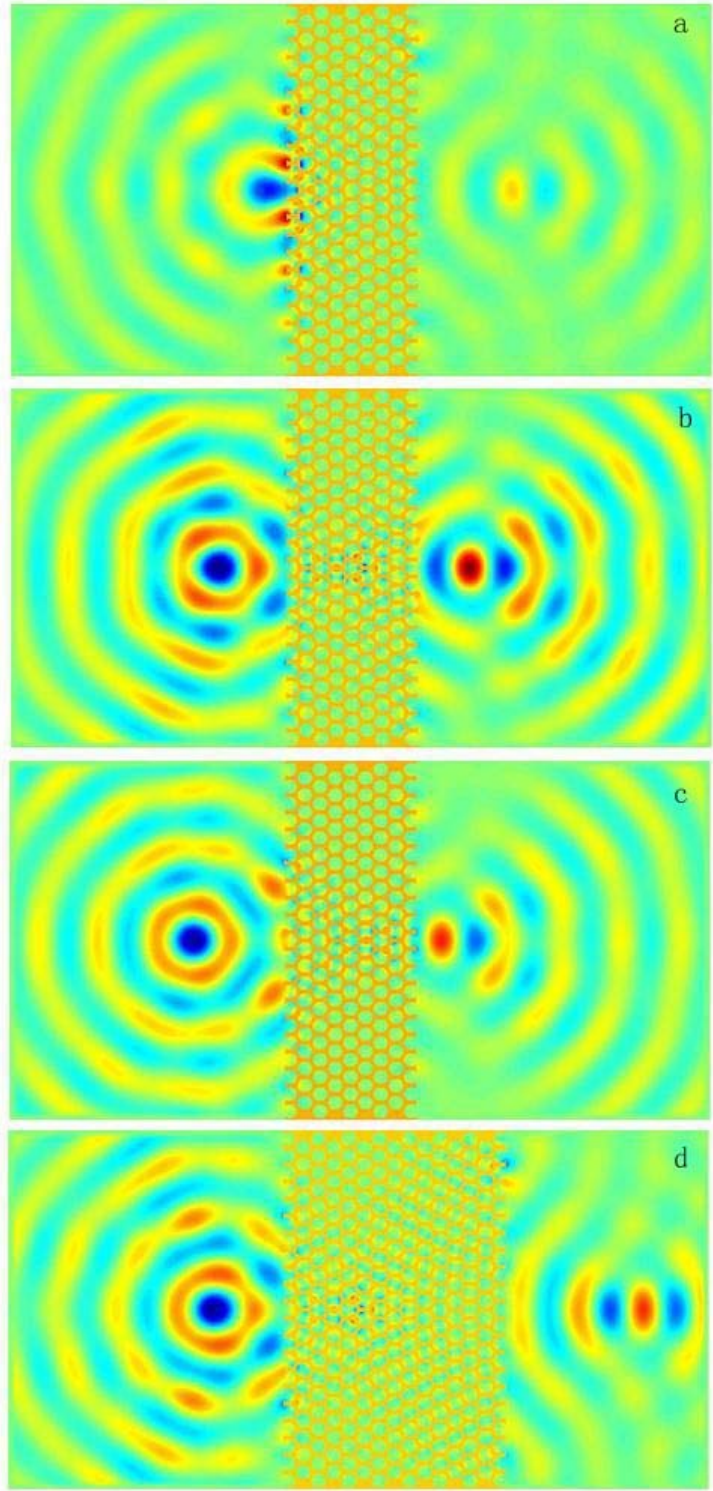

**Fig. 2**

**X. Wang et al.**

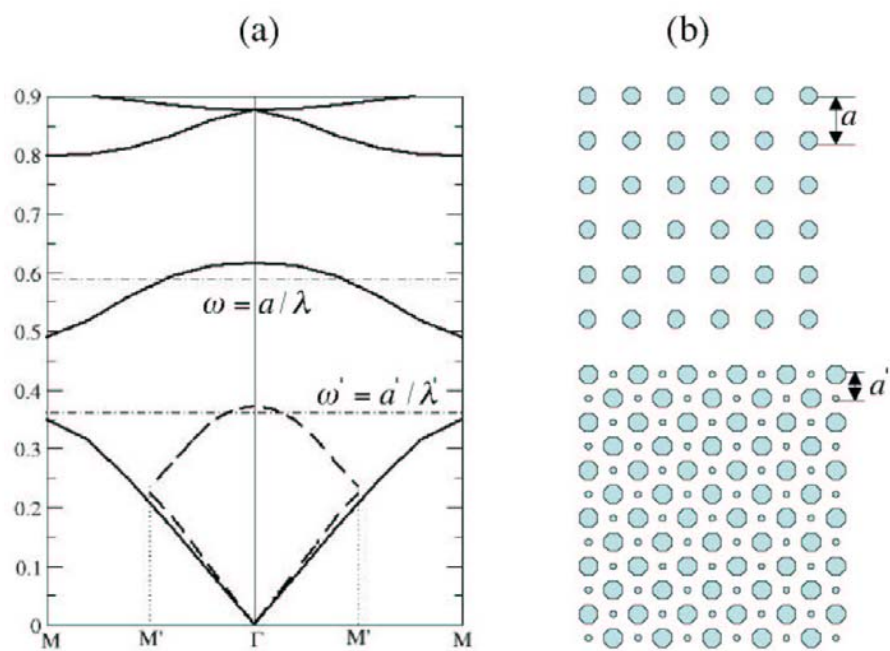

**Fig. 3**

**X. Wang et al.**

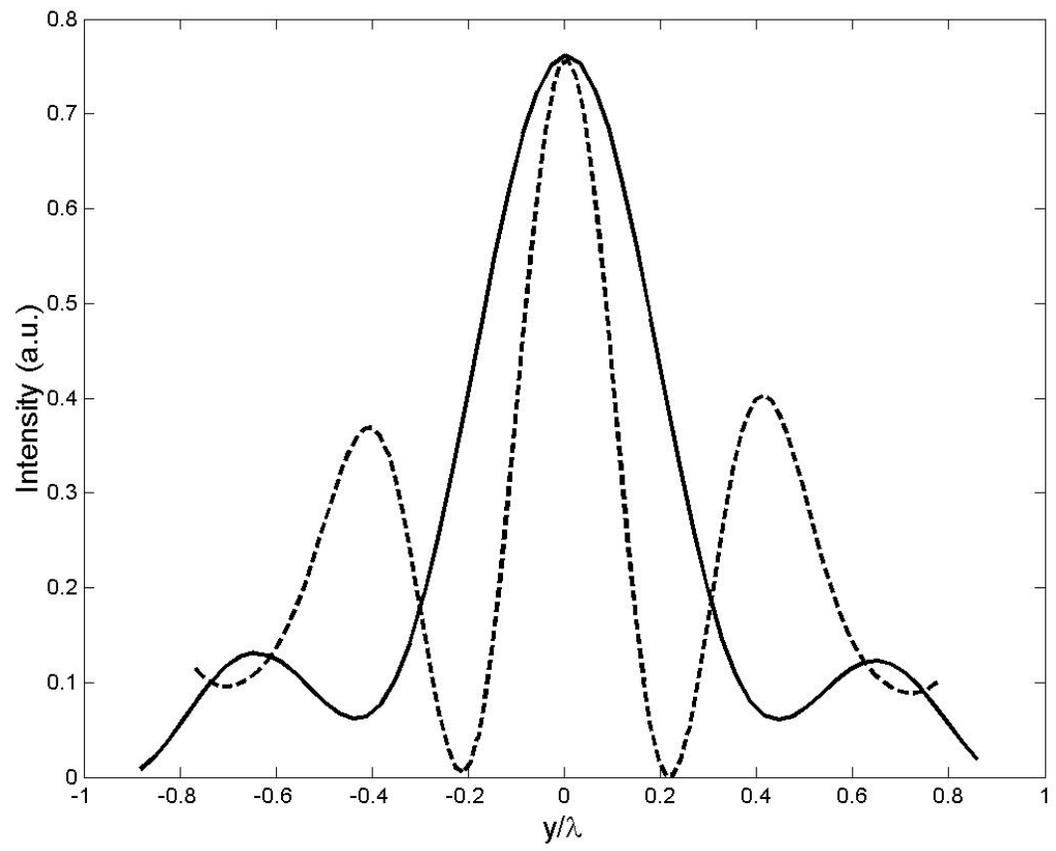

**Fig. 4**

**X. Wang et al.**

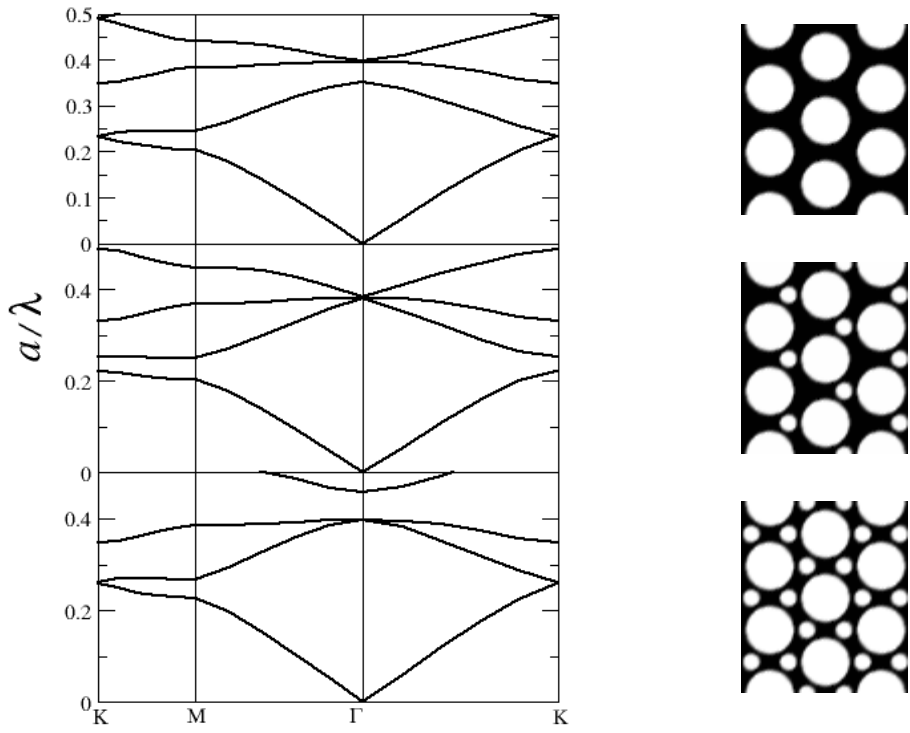

**Fig. 5**

**X. Wang et al.**

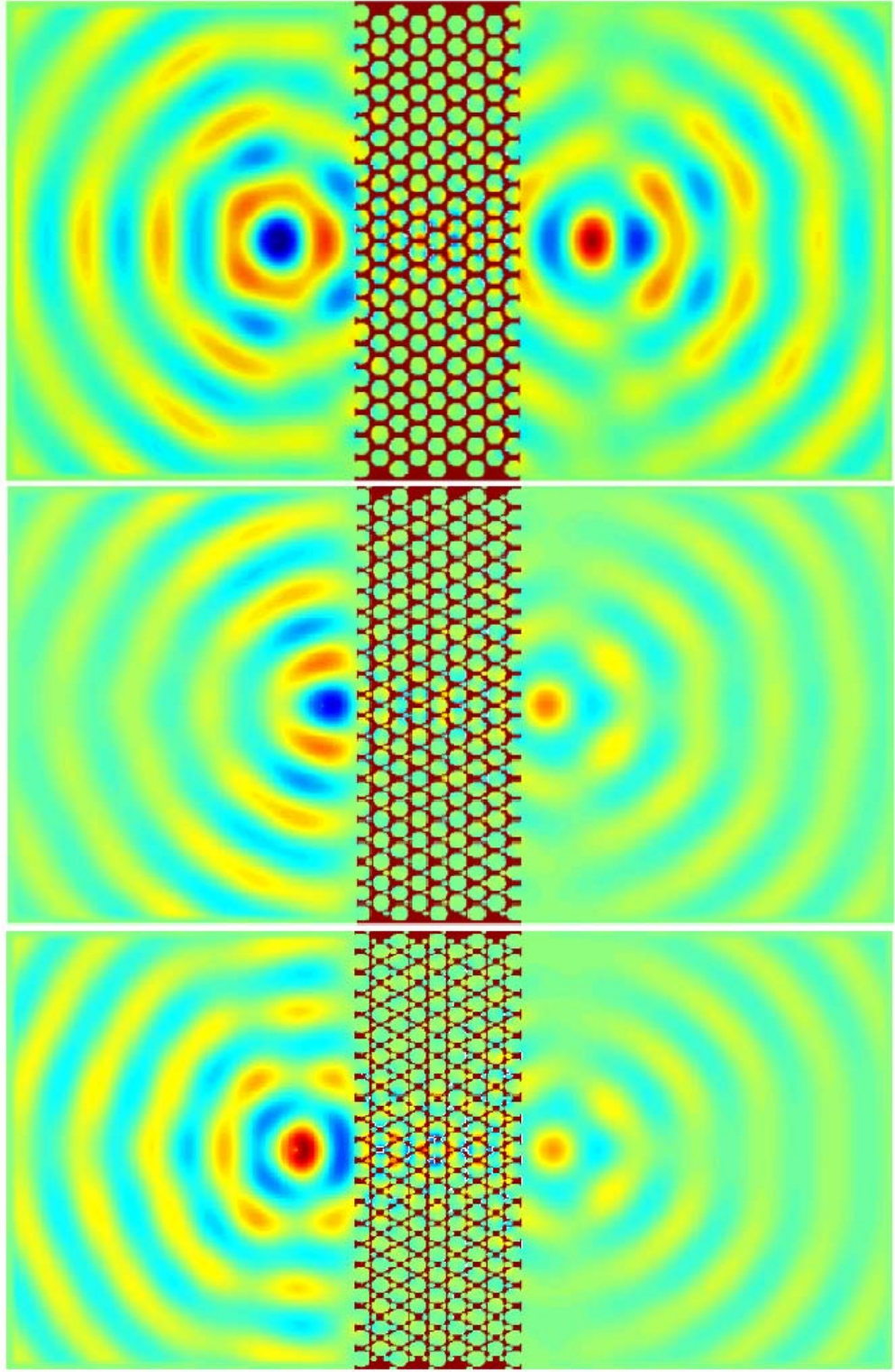

**Fig. 6**

**X. Wang et al.**